
\input phyzzx.tex

\def\prl{Phys. Rev. Lett. }

\overfullrule=0pt
\tolerance=5000
\overfullrule=0pt
\twelvepoint

\REF\HALD{F.D.M. Haldane, \prl {\bf 51}, 605 (1983).}
\REF\HALP{B.I. Halperin, Phys. Rev. Lett. {\bf 52}, 1583 (1984).}
\REF\BW {B. Blok, X.G. Wen, Phys. Rev. {\bf B42}, 8145 (1990); ibid. {\bf
43}, 8337 (1991).}
\REF\REA { N. Read, Phys. Rev. Lett. {\bf 65}, 1502 (1990).}
\REF\GREAD{G. Moore and N. Read, Nucl. Phys. {\bf 360B}(1991)362.}
\REF\PERIN{B.I. Halperin,  Helv Phys Acta {\bf 56}, 75 (1983).}
\REF\HALR{F.D.M. Haldane and E.H. Rezayi, Phys. Rev. Lett.{\bf 60},
956 (1988); Erratum:{\bf 60}, 1886 (1988).}
\REF\GP {R. Prange and S.M. Girvin., {\it The Quantum Hall Effect}
(Springer-Verlag, New York, Heidelberg, 1990, 2nd ed).}
\REF\LINN{D. Li, Phys. Lett. {\bf A169}(1992)82;
{\it Anyons and Quantum Hall Effect on the Sphere},
SISSA/ISAS/129/91/EP.}
\REF\LAO{R.B. Laughlin, \prl {\bf 50}, 1395 (1983). }
\REF\FRAN{D. Yoshioka, A. MacDonald, S.M. Girvin Phys. Rev. {\bf B39},
1932 (1988); F. Wilczek, Phy. Rev. Lett. {\bf 69}, 132 (1992).}
\REF\ASW {D.P. Arovas, J.R. Schriffer, and F. Wilczek, Phys. Rev.
Lett. {\bf 53}, 722 (1984).}
\REF\WI{F. Wilczek, {\it Fractional Statistics and  Anyon Superconductivity}
(World Scientific, 1990).}
\REF\ZHANG{S.C. Zhang, Int. J. Mod. Phys. {\bf B6}, 25 (1992).}
\REF\WEN{X.G. Wen and A. Zee, Phys. Rev. Lett {\bf 69}, 953 (1992);
Erratum: {\bf 69}, 3000 (1992).}
\REF\KIV{S.L. Sondhi and S.A. Kivelson, Phys. Rev. {\bf B46}, 13319 (1992).}
\REF\COL{S. Coleman, ``The magnetic monopole fifty years later",
lecturers given at the 1981 ``International School of Subnuclear Physics"
(Ettore Majorana), Erice, Italy.}
\REF\ROB{R. Iengo and K. Lechner,
Phys. Rep. {\bf 213}, 179 (1992).}
\REF\FCC{S.M. Girvin, the appendix in [\GP].}
\REF\FRACT{R.B. Laughlin, ``Fractional Statistics in the Quantum Hall Effect",
in [\WI].}
\REF\REZY{E.H. Rezayi, Phys. Rev. {\bf B39}, 13541 (1989).}
\REF\KANE{D-H. Lee and C.L. Kane, Phys. Rev. Lett.{\bf 64}, 1313 (1990).}
\REF\STU{J Fr{\" o}lich and U.M. Studer. {\it Incompressible Quantum Fluids,
Gauge-Invariance, and Current Algebra}; {\it
$U(1)\times SU(2)$-Gauge Invariance
of Nonrelativistic Quantum Mechanics, and Generalized Hall Effects},
preprint.}
\REF\CHERN{S.M. Girvin and A. MacDonald, \prl {\bf 56}, 1252 (1987);
S.M. Girvin, chapter 10 in [\GP].}
\REF\ZHK {S.C. Zhang, T.H. Hansson, and S. Kivelson, Phys. Rev.
Lett. {\bf 62}, 82 (1989).}
\REF\RE {N. Read, Phys. Rev. Lett. {\bf 62}, 86 (1989).}
\REF\WQ{X.G. Wen and Q. Niu, Phys. Rev. {\bf B41}, 9377 (1990).}
\REF\ZEE{A. Zee, ITP preprint 91-100.}
\REF\HR{ F.D.M. Haldane and E.H. Rezayi, Phys. Rev. {\bf B31},  2529 (85).}
\REF\LID{D. Li, {\it Hierarchical Wave Functions of Fractional Quantum
Hall Effect on the Torus}, SISSA/ISAS/2/92/EP;
{\it Hierarchical Wave Functions and Fractional
Statistics in Fractional Quantum Hall Effect on the Torus},
SISSA/ISAS/58/92/EP. To appear in Int. J. Mod. Phys. {\bf B}.}
\REF\MUN{D. Mumford, {\it Tata Lectures on Theta I and II}
(Birkh{\"a}user, Boston, 1983).}

\date{}
\titlepage
\title{Hierarchical
wave function, Fock cyclic condition and spin-statistics relation
in the spin-singlet fractional quantum Hall effect}
\vglue-.25in
\author{Dingping Li\foot{email: LIDP@TSMI19.SISSA.IT}}
\address{ International School for Advanced Studies,
I-34014 Trieste, Italy}
\bigskip
\abstract{We construct the hierarchical wave function of
the spin-singlet fractional quantum Hall effect, which turns out to satisfy
Fock cyclic condition.  The spin-statistics relation
of the quasi-particles in the spin-singlet
fractional quantum Hall effect is also discussed.
Then we use particle-hole conjugation to check the wave function.}
\par \noindent


\endpage

\chapter{INTRODUCTION}

The hierarchical wave function of the spin-polarized
fractional quantum Hall  effect  (FQHE) has been
constructed by Haldane [\HALD] and Halperin [\HALP] and
it also has been subjected
to intensive studies in the last several years [\BW,\REA,\GREAD, etc.].
When the magnetic field is not strong enough, the electron spin  maybe
not be polarized. Halperin had proposed a class of state with half spins
reversed which are spin-singlet states [\PERIN] and
Haldane and Rezayi had proposed a spin-singlet state
at a filling factor $\nu ={1\over 2}$ [\HALR].
However the hierarchical
wave function for the spin-non-polarized case,
for example, spin-singlet FQHE (SFQHE),
is not still fully understood as it was
pointed out by Girvin [\GP]. In particular, it
is much more difficult to  obtain
the hierarchical wave function of SFQHE subjected to
the requirement of Fock cyclic condition (FCC) (we will discuss
it in section $4$).
So it remains an interesting problem. The task of the present paper is to
construct the hierarchical wave function based
on Halperin spin-singlet state. However we shall point out that
this hierarchical scheme does not include
Haldane and Rezayi state [\HALR].
\par
We will use the projective coordinate to construct hierarchical
wave function of SFQHE on the sphere. The projective coordinate has been used
to construct hierarchical wave function in the spin-polarized
FQHE on the sphere [\LINN]. And here
the  same notation will be used as that  in [\LINN].
\par
In the projective coordinates $x, y$,
the metric $g$ of the sphere is given by
$$g_{\alpha\beta}(x)={1\over(1+r^{\prime2})^2}
\left(\matrix{1&0\cr
0&1\cr}\right) \,  \, .\eqn\ablin$$
where   $r^{\prime2}={x^2+y^2\over 4 R^2}$, and
$R$ is the radius of the sphere.
For simplicity, we assume the radius of the sphere $R={1\over 2}$.
The hamiltonian of electron
in a magnetic monopole field is
(the hamiltonian with  Laplace-Beltrami operator ordering differs
from this  hamiltonian by a constant) then
$$H={2\over M_e}{(1+z \bar z)}^2(P_z-eA_{z})
(P_{\bar z}-eA_{\bar z}) \, \, .\eqn\bcslin$$
The magnetic monopole field $eA_{z}$ is
$$eA_{z}=-{\imath \phi \over 2}{\bar z \over
1+z \bar z} \, \,  ,\eqn \bctlin$$
where  $\phi$ is the magnetic flux in
the  unit of the fundamental flux and it is an integer.
$P_{z_i}$ and $P_{\bar z_i}$ in \bctlin  are the operators
$$P_{z_i}=-\imath \partial {z_i} \, \, ,P_{\bar z_i}=-\imath
\partial_{\bar z_i} \, \, . \eqn \bcttlin$$
We have put the Dirac singularity of the monopole field at $z=\infty$.
By solving the equation $(P_{\bar z}-eA_{\bar z})\Psi =0$,
the ground state of the electron is
$$\psi=z^k (1+z \bar z)^{-{\phi \over 2}}  \,\, ,\eqn\bculin$$
where $0 \leq k \leq \phi$ with $k$ being an integer
(in order that  the  ground state is normalizable).
The Laughlin wave function [\LAO] shall be
$$\Psi=\prod _{i<j} (z_i-z_j)^m \prod_i (1+z_i
\bar z_i)^{-{\phi \over 2}}  \,\, ,\eqn\bcv$$
where $m$ is an odd integer.
It is known that the state of FQHE
on the sphere and plane is non-degenerate due to the reason that, by
adding Coulomb interaction, the Landau degeneracy is lifted.
Suppose that the system still has rotational symmetry,
thus the ground must be rotationally invariant. Under the rotation,
the coordinate is transformed  as
$$z^\prime={az+b\over cz+d}\, \, , \left(\matrix{a&b\cr
c&d\cr}\right)\in SO(3) \, \, ,\eqn\aplinn$$
which is generated by the rotations around three axes
$$\eqalign{&R_x={1\over \sqrt 2}\pmatrix{
\left(1+\cos\alpha\right)^{1\over 2}&
\imath \left(1-\cos\alpha\right)^{1\over 2}\cr
\imath \left(1-\cos\alpha\right)^{1\over 2}&
\left(1+\cos\alpha\right)^{1\over 2}
\cr} \, \, , \cr
&R_y={1\over \sqrt 2}\pmatrix{
\left(1-\cos\beta\right)^{1\over 2}&
\left(1-\cos\beta\right)^{1\over 2}\cr
-\left(1-\cos\beta\right)^{1\over 2}&
\left(1-\cos\beta\right)^{1\over 2}\cr}\, \, ,\cr
&R_z=\pmatrix{
\exp {\imath \gamma\over2}&0\cr
o&\exp {-\imath \gamma\over2}\cr} \,  \, .\cr}   \eqn\aqlinn$$
Under the rotation $ z^\prime={az+b\over
cz+d}$, the wave function is transformed into [\LINN]
$$R(a,b,c,d)\Psi (z_i)=\prod_i ({\bar c \bar z_i+\bar z_i
\over cz_i+d})^{-\phi \over 2} \Psi ({az_i+b\over cz_i+d})
\,  \, ,\eqn\bcwlin$$
where $R$ is the corresponding quantum operator of the rotation.
The  transformation of
$z_i-z_j$ and $1+z{\bar z}$  will be:
$$\eqalign{&z_i^\prime -z_j^\prime
={z_i-z_j\over \left(cz_i+d\right)
\left(cz_j+d\right)} \, ,\cr
&1+z^\prime \bar z^\prime ={1+z\bar z \over \left(cz+d\right)
\left(\bar c \bar z +\bar d \right)}\, , \cr}   \eqn\arlinn$$
and
$$d_{ij}={z_i-z_j \over
(1+z_i \bar z_i)^{1\over 2}(1+z_j \bar z_j)^{1\over 2}} \, ,
\eqn\defimm$$
will be transformed as
$$d_{ij}^{\prime}=({{\overline {cz_i+
d}}\over  cz_i+d})^{1\over 2} ({{\overline {cz_j+d}}\over  cz_j+d}
)^{1\over 2}d_{ij} \, \, . \eqn\xxxy$$
Implementing the wave function being
rotationally invariant, that is, $R\Psi=\Psi$,
one gets
$$\phi=m(N-1)  \,\, ,\eqn\bcxlin$$
where $N$ is the number of the electrons. Then
the wave function \bcv is equal to $\prod_{i<j}d_{ij}^m$ [\HALD].
\par
The hierarchical wave function can be constructed
as follows (for the case on the plane, see [\BW]).
The {\it normalized} wave function in the presence of quasi-particles at
$z_{\alpha}^{\prime}$ and at the filling $\nu ={1\over m}$ is
$$\Psi_e=\prod_{i<j} d^m_{ij} \prod_{i\alpha} d_{i\alpha}
\prod_{\alpha <\beta}{d_{\alpha\beta}}^{1\over m} \,\, ,\eqn\hholin$$
The Laughlin type wave function of the quasi-particles is
$$\Psi_q=\prod_{\alpha <\beta}{\bar d_{\alpha\beta}}^{1\over m}{({\bar
d_{\alpha\beta}})}^p \, \, ,\eqn\hhplin$$
where $p$ is a positive even integers.
The hierarchical wave function of the electrons is given by
$$\Psi (z_i)=\int \prod_{\alpha} dv_{\alpha}
\Psi_e(z_i, z_{\alpha}^{\prime})\Psi_q
(z_{\alpha}^{\prime})   \, \, ,\eqn\hhnlin$$
where $dv_{\alpha}={d_{z_{\alpha}^{\prime}}^2
 \over (1+z_{\alpha}^{\prime} \bar z_{\alpha}^{\prime})^2}$
is rotationally  invariant measure on the sphere.
Requiring   $\Psi (z_i)$ to be rotationally invariant, $R\Psi (z_i)=R$,
we get the relation
$$\eqalign{&m(N_e-1)+N_q=\phi \, ,\cr
&N_e-p(N_q-1)=0 \, \, ,\cr}    \eqn\relation$$
where $N_e$ is the electron number and $N_q$
is the quasi-particle number.
{}From \relation , one can derive  the filling
$$\nu={1\over m+{1\over p}} \, \, . \eqn\retion$$
By considering the excited state of \hhplin
and let those quasi-particles of the quasi-particles be condensed,
we can get the next hierarchical state.
By proceeding in this way, the general
hierarchical state can be constructed.
\par
We shall generalize the above method to construct the hierarchical
wave function of SFQHE. In the next section we review some basic facts
about FQHE with the layered structure, which is useful for the discussions
in the following sections. In section $3$, we discuss the
spin-statistics relation of the quasi-particles in the
layered FQHE. In section $4$, we construct an hierarchical
wave function of SFQHE and prove that the wave
function satisfies FCC. In section $5$, we try to give a physical explanation
of the wave function obtained in section $4$.
In section $6$, we apply the particle-hole conjugation
operation to the wave function. Finally in section $7$, we make
the conclusion of the paper.

\chapter{THE LAYERED FQHE }

Halperin had proposed some states
with half spin reversed [\PERIN],
$$\Psi_{mmn}=\prod_{i<j}^{N}\prod_{k,l}^{N} [d(z_i,z_j)]^m
[d(w_i,w_j)]^m[d(z_k,w_l)]^n  \, ,  \eqn\aaa$$
where $m$ is an odd integer,
$z_i$ are the coordinates of the up-spin electrons and
$w_i$ are the coordinates of the down-spin electrons.
We can also interpret $z_i$ as the coordinates
of the up-layer electrons and
$w_i$ as the coordinates of the
down-layer electrons in a double layered
FQHE.  Now We would like to discuss a  more general
type wave function [\FRAN],
$$\Psi_{m_1,m_2,n}=\prod [d(z_i,z_j)]^{m_1}
[d(w_i,w_j)]^{m_2}[d(z_k,w_l)]^n  \, .\eqn\aab$$
Because the up-layer electrons $z_i$ can be distinguished
from the down-layer electrons $w_i$,
so the wave function does not need to be
completely anti-symmetrized.
$\Psi_{m_1,m_2,n}$ is rotationally invariant [\HALD,\LINN],
and we have
$$m_1(N_1-1)+nN_2=nN_1+m_2(N_2-1) =\phi \, , \eqn\aac$$
because all electrons are exposed to the same magnetic field
and in the lowest Landau level. $N_1$ ($N_2)$ is the number of
the up (down)-layer electrons and $\phi$ is the magnetic flux.
According to \aac, the filling is ($N_1, N_2, \phi$ etc.
are always assumed to be much larger than $1$)
$$\nu={N_1+N_2 \over \phi}={m_1+m_2-2n\over m_1m_2-n^2}\, .\eqn\aad$$
When $m_1=m_2=m$, the filling is then $2\over m+n$.
Now we introduce a two dimension lattice with bases
$$e_i \cdot e_j=\Lambda_{i,j}=\pmatrix{
m_1&n\cr
n&m_2\cr} \, . \eqn\aae$$
The bases of the inverse lattice
is defined by $e^{\ast}_i\cdot e_j=\delta_{i,j}$,
and thus we have
$$e^{\ast}_i\cdot e^{\ast}_j
=\Lambda^{-1}_{i,j}={1\over m_1m_2-n^2}\pmatrix{
m_2&-n\cr -n&m_1\cr} \, . \eqn\aaf$$
The wave function $\Psi_{m_1,m_2,n}$ can be written now as
$$\Psi_{m_1,m_2,n}=\prod [d(z_i,z_j)]^{e_1\cdot e_1}
[d(w_i,w_j)]^{e_2\cdot e_2}[d(z_k,w_l)]^{e_1\cdot e_2}  \, .\eqn\aag$$
The wave function with quasi-particles at
$z^{\prime}_{\alpha}$ and $w^{\prime}_{\alpha}$ is
$$\eqalign{\Psi_{m_1,m_2,n}(z^{\prime}_{\alpha},w^{\prime}_{\alpha})&=
\prod d(z_i-z^{\prime}_{\alpha})
d(w_i-w^{\prime}_{\alpha})\Psi_{m_1,m_2,n} \cr &
=\prod d(z_i-z^{\prime}_{\alpha})^{e_1 \cdot e^{\ast}_1}
d(w_i-w^{\prime}_{\alpha})^{e_2 \cdot
e^{\ast}_2}\Psi_{m_1,m_2,n} \, .\cr} \eqn\aah$$
The equation \aac now becomes
$$\eqalign{&m_1(N_1-1)+nN_2+N^{\prime}_1=\phi \, , \cr &
nN_1+m_2(N_2-1)+N^{\prime}_2 =\phi \,  , \cr } \eqn\naal$$
where $N_1^{\prime}$ ($N_2^{\prime}$) is
the number of the up(down)-layer quasi-particles.
The plasma charge of the electron $z_i$ ($w_i$) is
$e_1$ ($e_2$) and  the plasma charge of the
quasi-particle $z_{\alpha}^{\prime}$ ($w_{\alpha}^{\prime}$)
is  $e^{\ast}_1$ ($e^{\ast}_2$).
The normalized wave function can be obtained
by using plasma analogue on the sphere,
$$\eqalign{\Psi_{m_1,m_2,n}(z^{\prime}_{\alpha},w^{\prime}_{\alpha})_{nor}
=&\prod_{i,\alpha}
d(z_i-z^{\prime}_{\alpha})^{e_1 \cdot e^{\ast}_1}
d(w_i-w^{\prime}_{\alpha})^{e_2 \cdot e^{\ast}_2}
d(z^{\prime}_{\alpha}-z^{\prime}_{\beta})^{e^{\ast}_1 \cdot e^{\ast}_1}
\cr & \times
d(w^{\prime}_{\alpha}-w^{\prime}_{\beta})^{e^{\ast}_2 \cdot e^{\ast}_2}
d(z^{\prime}_{\alpha}-w^{\prime}_{\beta})^{e^{\ast}_1 \cdot e^{\ast}_2}
\Psi_{m_1,m_2,n}\, .\cr} \eqn\aah$$
The normalization constant of the above wave function
will be independent on the coordinates of the
quasi-particles in the limit of the
quasi-particles being quite far away from each other.
The statistics parameter $\theta_{ij}$
(when exchanging two kinds of particles $i$ and $j$,
we will get a phase $e^{i\theta_{ij} \pi}$)
of the quasi-particle can be read from the
the wave function \aah,
$${\theta_{ij}}=-e^{\ast}_i\cdot e^{\ast}_j=
-\Lambda^{-1}  \,  . \eqn\aai$$
The electric charge of the quasi-particle $z^{\prime}_{\alpha}$
is  $\Lambda^{-1}_{1,1}+\Lambda^{-1}_{1,2}={{m_2-n}\over m_1m_2-n^2}$
and the charge of the quasi-particle $w^{\prime}_{\alpha}$
is $\Lambda^{-1}_{2,2}+\Lambda^{-1}_{2,1}={{m_1-n}\over m_1m_2-n^2}$,
where the electron charge is assumed to be $-1$
(the above results can be derived by using Berry phase method
[\ASW] or the article by D.P. Arovas in [\WI].
however see also the  next section).

\chapter{HIERARCHICAL WAVE FUNCTION AND SPIN-STATISTICS RELATION IN
THE LAYERED FQHE}

The hierarchical wave function
can be obtained when the quasi-particles
are condensed. We can analyze the hierarchical wave function to
obtain the spin of the quasi-particle [\LINN]. The idea is that,
from the hierarchical wave function, we can obtain the hamiltonian
of the quasi-particles and then the spin of the quasi-particle
by analyzing the hamiltonian.
The spin of the quasi-particles
can also be obtained by calculating  Berry phase when
the quasi-particle  moves in a closed path [\LINN].
\par
The hamiltonian of the quasi-particles
can be obtained by using the fact that
the suggested wave function of quasi-particles,
which is Laughlin type [\HALP],
is the  ground state of the hamiltonian, or by using Berry phase method
[\BW].
We shall remark that, the lagrangian of
the quasi-particles are described by vortex (center coordinate) dynamics,
and the lagrangian of vortices (quasi-particles)
does not contain any mass term [\ZHANG].
The Hilbert space of the hamiltonian which we will
derive in the following shall be restricted to ground state.
The ground state of the following hamiltonian is the same as the
one  obtained by analyzing
the vortex dynamics theory of the quasi-particles.
So to be rigorous, we shall
proceed our discussion from the beginning based
on vortex dynamics theory.
\par
The problem about the spin of the quasi-particle
has also been addressed in [\WEN,\KIV].
The result about the spin of the quasi-particle in
[\LINN] agrees with the one in [\WEN].
However there are some differences in the definition of the spin
between [\LINN] and [\WEN], which we explain later.
$s$ in [\LINN] corresponds to $S_{total}$ [\WEN].
The reference
[\WEN] has also calculated the spin of the quasi-particle in
general hierarchical state and multilayered FQHE state (which
we can also use the method described in [\LINN] to calculate).
The spin-statistics relation
of the quasi-particle in general hierarchical state and multilayered
FQHE state usually is not standard one [\WEN],
even it is found that the quasi-particle
in Laughlin state (with filling as $1\over m$)
has standard spin-statistics relation [\LINN].
We are aware that the spin of the quasi-particle in the Laughlin state
calculated in [\KIV] is different from the one in [\LINN,\WEN].
\par
Let us consider
the special case $m_1=m_2$ of the last section for simplicity.
Then we have
$$e^{\ast}_i\cdot e^{\ast}_j={1\over m^2-n^2}\pmatrix{
m&-n\cr
-n&m\cr} \,  . \eqn\aainx$$
The  wave function of
the condensed quasi-particles  in the singular gauge
is a Laughlin type wave function,
and according to \aai,    it shall be
$$\eqalign{{\tilde {\Psi}^q_p} =&d(z^{\prime}_{\alpha}-
z^{\prime}_{\beta})^{e^{\ast}_1 \cdot e^{\ast}_1}
d(w^{\prime}_{\alpha}-w^{\prime}_{\beta})^{e^{\ast}_2 \cdot e^{\ast}_2}
d(z^{\prime}_{\alpha}-w^{\prime}_{\beta})^{e^{\ast}_1 \cdot e^{\ast}_2}\cr &
\times [d(z^{\prime}_{\alpha}-
z^{\prime}_{\beta})d(w^{\prime}_{\alpha}-w^{\prime}_{\beta})
d(z^{\prime}_{\alpha}-w^{\prime}_{\beta})]^p  \, .\cr} \eqn\aaj$$
Hence the hierarchical wave function is
$$\Psi_{mmn,p}=\int \prod dv_{\alpha}
\Psi_{mmn}(z_i,w_i, z^{\prime}_{\alpha}, w^{\prime}_{\alpha})\Psi^q_p
(z^{\prime}_{\alpha},w^{\prime}_{\alpha})   \,  ,\eqn\aak$$
where $\Psi_{mmn}(z_i,w_i, z^{\prime}_{\alpha}, w^{\prime}_{\alpha})$
now is the normalized wave function given by \aah and
$dv_{\alpha}$ are the rotationally invariant volume measures
of the quasi-particles.
By imposing the rotationally invariant condition on the wave function
$\Psi_{mmn,p}$, we can obtain the relation
$$\eqalign{&m(N_1-1)+nN_2+N^{\prime}_1=\phi \, , \cr &
nN_1+m(N_2-1)+N^{\prime}_2 =\phi \, ,  \cr &
N_1-p(N^{\prime}_1-1)-pN^{\prime}_2=0 \, , \cr &
N_2-p(N^{\prime}_2-1)-pN^{\prime}_1=0 \, . \cr } \eqn\aal$$
The first two equations in \aal are equations in \naal.
{}From \aal, we get the filling $\nu$,
$$\nu ={2\over m+n+{1\over 2p}} \, \, . \eqn\nnaal$$
To discuss the hamiltonian of the quasi-particles, we will use
the quasi-particle wave function in non-singular gauge,
$$\eqalign{{\Psi}^{\prime}_p =&|d(z^{\prime}_{\alpha}-
z^{\prime}_{\beta})^{e^{\ast}_1 \cdot e^{\ast}_1}
d(w^{\prime}_{\alpha}-w^{\prime}_{\beta})^{e^{\ast}_2 \cdot e^{\ast}_2}
d(z^{\prime}_{\alpha}-w^{\prime}_{\beta})^{e^{\ast}_1 \cdot e^{\ast}_2}|\cr &
\times [{\bar d}(z^{\prime}_{\alpha}-
z^{\prime}_{\beta}){\bar d}(w^{\prime}_{\alpha}-w^{\prime}_{\beta})
{\bar d}(z^{\prime}_{\alpha}-w^{\prime}_{\beta})]^p  \, ,\cr} \eqn\aajn$$
The hamiltonian which has  ${\Psi}^{\prime}_p$
as the ground state is
$$\eqalign{H=&{2\over M}\sum \left(1+z^{\prime}_i\bar
z^{\prime}_i\right)^2 (P_{\bar z^{\prime}_i} -A_{\bar z^{\prime}_i})
(P_{z^{\prime}_i}-A_{z^{\prime}_i})+ \cr &
\left(1+w^{\prime}_i\bar
w^{\prime}_i\right)^2 (P_{\bar w^{\prime}_i} -A_{\bar w^{\prime}_i})
(P_{w^{\prime}_i}-A_{w^{\prime}_i})  \,  , \cr}\eqn\aam$$
where
$$\eqalign{A_{z^{\prime}_i}=&{-im\over 2(m^2-n^2)}\sum_{j\not=i}
{1\over z^{\prime}_i-z^{\prime}_j}+{in\over 2(m^2-n^2)}\sum_{i,j}
{1\over z^{\prime}_i-w^{\prime}_j}+ \cr &
{i\over 2}{m(m-n-1)\over (m-n)(m+n)}
{{\bar z^{\prime}_i}\over 1+z^{\prime}_i\bar z^{\prime}_i}
+{i\phi \over 2(m+n)} {{\bar z^{\prime}_i}\over 1
+z^{\prime}_i\bar z^{\prime}_i} \,  ,\cr} \eqn\aan$$
and
$$\eqalign{A_{w^{\prime}_i}=&{-im\over 2(m^2-n^2)}\sum_{j\not=i}
{1\over w^{\prime}_i-w^{\prime}_j}+{in\over 2(m^2-n^2)}\sum_{i,j}
{1\over w^{\prime}_i-z^{\prime}_j}+ \cr &
{i\over 2}{m(m-n-1)\over (m-n)(m+n)}
{{\bar w^{\prime}_i}\over 1+w^{\prime}_i\bar w^{\prime}_i}
+{i\phi \over 2(m+n)} {{\bar w^{\prime}_i}\over 1
+w^{\prime}_i\bar w^{\prime}_i} \,  .\cr} \eqn\naan$$
we can check that, by using
the relation \aal ,
$P_{z^{\prime}_i}-A_{z^{\prime}_i}$ or
$P_{w^{\prime}_i}-A_{w^{\prime}_i}$ acting on the wave function
${\Psi}^{\prime}_p$ is zero.
The lagrangian of the quasi-particles
is (for the case of disc geometry, see [\ZHANG]),
$$L=\sum  A_{z^{\prime}_i} {dz^{\prime}_i \over dt}+
A_{\bar z^{\prime}_i} {d \bar z^{\prime}_i \over dt}+
A_{w^{\prime}_i} {dw^{\prime}_i \over dt}+
A_{\bar w^{\prime}_i} {d \bar w^{\prime}_i \over dt}. \eqn\lagran$$
{}From the lagrangian, we  use Noether theorem
to derive the angular momenta of the quasi-particle.
Then from the angular
momenta, we can get the spin of the quasi-particle.
The first and second terms in the right of the equations
\aan and \naan tell us that
the quasi-particles satisfy fractional statistics.
The last terms in  \aan and \naan represent the interaction
between the quasi-particles and magnetic field
(so the electric charge of the quasi-particle
is $1\over m+n$).
The statistics parameters and
the electric  charge of the quasi-particle
given by \aan and \naan
are consistent with  the ones given by the last section.
\par
The spin of the particle will be changed
by the presence of the magnetic monopole
field or other particles with monopole
charges [\COL]. For example, if a electron interacts with a magnetic
monopole of odd integer flux, the spin of the electron will
be an integer instead of $1\over 2$.
The spin we would like to discuss is
the intrinsic spin which shall not depend on the presence
of the applied magnetic monopole field
or other particles with monopole charges.
\par
Let us consider the spin of quasi-particle $z^{\prime}_i$.
By calculating
Noether currents of rotational invariance of the lagrangian
([\LINN] or the chapter $3$ in [\ROB]),
we find that the terms  in \lagran, for example,
${-im\over 2(m^2-n^2)}
{1\over z^{\prime}_i-z^{\prime}_j}{dz^{\prime}_i \over dt}$ and
${in\over 2(m^2-n^2)}
{1\over z^{\prime}_i-w^{\prime}_j}$ (and their complex conjugate)
will contribute to spin of the quasi-particle.
Also the interaction between the quasi-particle and magnetic field,
which is described by the term  in \lagran as
${i\phi \over 2(m+n)} {{\bar z^{\prime}_i}\over 1
+z^{\prime}_i\bar z^{\prime}_i}{dz^{\prime}_i \over dt}$
(and its complex conjugate)
will contribute to spin of the quasi-particle.
However there is another term in \lagran,
${i\over 2}{m(m-n-1)\over (m-n)(m+n)}
{{\bar z^{\prime}_i}\over 1+z^{\prime}_i\bar z^{\prime}_i}
{dz^{\prime}_i \over dt}$ (and its complex conjugate),
which represents the interaction between the
quasi-particle and
a monopole field with flux ${m(m-n-1)\over (m-n)(m+n)}$.
This term
is independent on the presence of the
applied magnetic monopole field or the presence of other
quasi-particles. So its  contribution to the spin of
the quasi-particle is intrinsic.
The contribution to the spin is
${1\over 2}{m(m-n-1)\over (m-n)(m+n)}$.
Thus  we identify the intrinsic spin (from this time on,
we will just simply call
intrinsic spin as spin)
of the quasi-particle as
$$s={m(m-n-1)\over 2(m-n)(m+n)} \, .\eqn\aaomad$$
This result can also be obtained
by using the formula for $S_{total}$ in [\WEN].
\par
When $m=1, n=0$, up(down)-layer FQHE is an integer
quantum Hall effect with the
filling as $1$.
The quasi-particles are
electrons or the holes of the electrons. From \aaomad,
$s$ will be equal to zero when $m=1, n=0$.
However one should expect that the electron spin is not $0$, but
$1\over 2$. What is the reason for the deficit of $1\over 2$?
In the layered FQHE,
the Pauli spin of the up(down)-layer electron   is
polarized. So only one component of the electron is taken into
account and the Pauli spin $1\over 2$
usually is forgotten for this reason.
The wave function of the electron in the up-layer
is $\psi_{ui}(e)$ where $i$ is Pauli spin index
($i=1 \, , 2$). In the polarized case, one component is zero, for example,
$\psi_{u2}(e)=0$.
The same reasoning shall also be applied to the quasi-particle wave function.
The up-layer quasi-particle wave function
is $\psi_{ui}(q)$. In the polarized case,
$\psi_{u2}(q)=0$.
Thus We shall include the Pauli spin to the (intrinsic)
spin  of the quasi-particle and now the spin will
be equal to $s^t=s+{1\over 2}$. This definition of
the spin of the quasi-particle is consistent
with the fact that the electron spin is $1\over 2$.
\par
The generalized standard spin-statistics relation shall be
$$s^t={\theta \over 2}+integer \, . \eqn\stand$$
In the present problem, $\theta$ equals to
$-m\over m^2-n^2$ ($\theta$ corresponds to $\theta_{11}$
in the last section) and $s^t=s+{1\over 2}$ is given by the equation
\aaomad .
However  the relation \stand
is not  satisfied in this case.
\par
When $m=n+1$, the wave  function  $\Psi_{mmn}$ satisfies
Fock cyclic condition (we will discuss it in the next section). Thus
$\Psi_{n+1,n+1,n}$ can  be used to describe the
un-layered spin-singlet FQHE (SFQHE) [\FCC].
When $m=n+1$, $s^t$ equals to ${1\over 2}$.
However it is not clear to us that the definition of
the quasi-particle spin by the equation \stand
is suitable for the quasi-particle spin in
the spin-singlet FQHE or not.

\chapter{HIERARCHICAL WAVE FUNCTION WITH FOCK CYCLIC CONSTRAINT}

For FQHE with half spins reversed and without
layered structure,
all electrons are identical and
the wave function needs to
be completely anti-symmetrized. The wave function is [\FCC]
$$\Phi_{mmn}=\sum_P{(-1)^p \over (2N)!}
\Psi_{mmn}(z_{P(1)}, \cdot \cdots ,z_{P(N)};
z_{P(1+N)}, \cdots, z_{P(2N)})\Psi_s(P) \, , \eqn\aap$$
where $z_{i+N}=w_i$, $P$ is permutation operator,  $p$ is the parity of
the permutation $P$ and $\Psi_s(P)$ is the spin function
$$\Psi_s(P)=(\alpha_{P(1)} , \cdots , \alpha_{P(N)};
\beta_{P(N+1)} , \cdots , \beta_{P(2N)}) \, , \eqn\spin$$
where $\alpha$ and $\beta$ represent the spin-up and spin-down states.
$\Phi_{mmn}$ is the eigenstate of $S^z=\sum_i^{2N}S^z_i=0$.
However it may not be the eigenstate
of $S^2=0$ (if $S^2=0$, then $S^z$ must be zero).
Now the operator of the rotation is  given by
$$R^t(a,b,c,d)=R^s(a,b,c,d)R(a,b,c,d) \, , \eqn\roation$$
where $R(a,b,c,d)$ is defined as that in \bcwlin and $R^s(a,b,c,d)$
is the rotational operator on spin function.
Hence if the wave function \aap is rotationally invariant,
that is $R^t\Phi_{mmn}=\Phi_{mmn}$,
we  may require
$$R\Phi_{mmn}=\Phi_{mmn} \, , R^s\Phi_{mmn}=\Phi_{mmn}
\, . \eqn\center$$
The condition $R^s\Phi_{mmn}=\Phi_{mmn}$
is equal to the condition $S^2=0$.
If we require $S^2\Psi_{mmn}=0$,
then we  obtain Fock cyclic condition (FCC) on the
wave function $\Psi_{mmn}$ [\FCC].  FCC is a condition given by
$$E_{z_i}\Psi= \sum_j e(z_i, w_j)\Psi=\Psi \, , \eqn\aaq$$
where $e(z_i, w_j)$ is the operator
which exchanges the coordinates  $z_i$ and $w_j$ of the function.
If $\Psi=\Psi_{mmn}$, then  $m=n+1$ is the only solution of  FCC.
It is well-known fact that
$\prod_{i<j}(z_i-z_j)(w_i-w_j)$
satisfies FCC \aaq . Then we can easily
show that $\prod_{i<j}d(z_i,z_j)d(w_i,w_j)$
satisfies FCC and so does $\Psi_{n+1,n+1,n}$.
One interesting problem is how to construct
the hierarchical wave function $\Phi$ on which $S^2$ is $0$.
We may first construct the hierarchical
wave function $\Psi$ based on the parent state
$\Psi_{n+1,n+1,n}$ by using the construction
discussed in the last section (now we must have
$N_1=N^{\prime}_1$ in order to have a rotationally invariant state),
then  we wish that it fulfills FCC. We will show that it is indeed so!
Following the last section, the general hierarchical wave function of
$\Psi_{n+1,n+1,n}$ is
$$\eqalign{\Psi(z_i(1),w_i(1))=&\int \prod dv_q
\Psi_1(z_i(1),w_i(1);z_i(2),w_i(2)) \times \cr &
\Psi_2(z_i(2),w_i(2);z_i(3),w_i(3))   \times \cdots \cr &
\Psi_{l-1}(z_i(l-1),w_i(l-1);z_i(l),w_i(l))
\Psi_1(z_i(l),w_i(l)) \, .\cr} \eqn\aar$$
$z_i(1)=z_i$ ($w_i(1)=w_i$) are up(down)-spin
electron coordinates and $z_i(k)$ ($w_i(k)$)
are  up(down)-spin quasi-particle coordinates of $k^{th}$ hierarchy.
$\Psi_1(z_i(1),w_i(1);z_i(2),w_i(2))$ is
the normalized electron wave functions in the presence
of quasi-particles at $z_i(2),w_i(2)$ and
$\Psi_k(z_i(k),w_i(k);z_i(k+1),w_i(k+1))$ is the
normalized wave function of
quasi-particles in $k^{th}$ hierarchy
in the presence of the next ($(k+1)^{th}$)
hierarchical quasi-particles.
We mean that the quasi-particles in the second
hierarchy is the quasi-particles of the electrons, etc..
The index  in $z_i(k)$ ($w_i(k))$  ranges over
$1\leq i\leq N_k$ ($1\leq i\leq N_k^{\prime}$).
$N_1$ ($N_1^{\prime}$) is the number of
the up(down)-spin electrons and $N_k$
($N_k^{\prime}$) is the number of
the up(down)-spin quasi-particles in $k^{th}$ hierarchy.
The integration in \aar is over all quasi-particles coordinates
(excluding the electron coordinates) and
the integral measure over every  quasi-particle coordinate is
rotationally invariant measure  on the sphere
(see the first section).
\par
Let us define
$$\eqalign{&F_k=\prod_{i<j}d(z_i(k),z_j(k))
d(w_i(k),w_j(k))\prod_{m,n} d(z_m(k),w_n(k)) \, ,\cr &
G_k=\prod_{i<j}d(z_i(k),z_j(k)) d(w_i(k),w_j(k)) \, ,\cr &
S_k=\prod_{i,j}d(z_i(k),z_j(k+1)) d(w_i(k),w_j(k+1)) \, .\cr} \eqn\aat$$
Then
the general hierarchical wave function $\Psi$ in \aar is given by
$$\eqalign{&\Psi_1=F^{p_1}_1G_1S_1(F_2)^{-p_1\over 2p_1+1}G_2
\, , \cr & \ldots \cr &
{\tilde \Psi_k}={(F_k)}^{s_k+p_k}
G_kS_k(F_{k+1})^{-(s_k+p_k)\over 2(s_k+p_k)+1}G_{k+1} \, ,\cr &
\ldots \cr &
{\tilde \Psi_l}={(F_l)}^{s_l+p_l}G_l \, . \cr}\eqn\aax$$
$p_1=n$, $p_l$ with $l>1$ are even integers  and
$${\tilde \Psi_k}=\cases {\Psi_k, &if $k=odd \, \,   integer$;
\cr {\bar \Psi_k}, &otherwise. \cr} \eqn\bbba$$
$s_k$ is given by the recursion relation
$$s_{k+1}=-{s_k+p_k \over 2(s_k+p_k)+1} \, ,
\eqn\aay$$
with $s_1=0$.
The statistics parameter of the condensed
quasi-particle in $k^{th}$ hierarchy is
$$\theta_k=(-1)^{k-1}(s_k+1) \, ,\eqn\aays$$
The charge of the condensed quasi-particle
in $k^{th}$ hierarchy is given by the recursion relation
$$e_{k+1}=-{e_k \over 2(s_k+p_k)+1} \, , \eqn\char$$
with the electron  charge $e_1=-1$.
\par
The wave function with $l$-hierarchies is characterized
by the $2l \times 2l$ matrix $\Lambda$
$$\Lambda =\pmatrix{I+p_1C&I&0&\ldots&0&0\cr
I&-p_2C&-I&0&\ldots&0\cr
0&-I&p_3C&I&0&\ldots\cr
\vdots&\vdots&\ddots&\ddots&\vdots&\vdots\cr
\vdots&\vdots&\ddots&\ddots&\vdots&\vdots\cr
0&\ldots&0&(-1)^{l-1}I&(-1)^{l-2}p_{l-1}C&(-1)^{l}I\cr
0&0&\ldots&0&(-1)^{l}I&(-1)^{(l-1)}p_lC\cr} \, , \eqn\aau$$
where $p_i$ positive even integers (except $p_1$ can be zero)
and $I, C$ are matrices,
$$I=\pmatrix{1&0\cr 0&1\cr}\, , C=\pmatrix{1&1\cr 1&1\cr}\, . \eqn\aav$$
In order that the wave function $\Phi$ be rotationally invariant, we
apply the first condition in \center and thus obtain the relation,
$$\sum_j \Lambda_{i,j} (H_j-\delta_{i,j})=\cases
{\phi, &if $i=1,2$; \cr 0, &otherwise, \cr}\eqn\aaww$$
where
$$H_{2i-1}=N_i, H_{2i}=N_i^{\prime} \, . \eqn\aaw$$
It is clear that $N_k=N_k^{\prime}$ in this case.
{}From \aaww we can derive filling as
$$\nu ={2\over \displaystyle 2p_1+1+
{\strut 1\over \displaystyle 2p_2+
{\strut 1\over \displaystyle \cdots +
{\strut 1\over \displaystyle 2p_l}}}}  \, .\eqn\fillingn$$
The second condition in \center is equal to Fock cyclic condition (FCC)
on $\Psi$. So  we  need to prove that the  wave function \aar
satisfies FCC.  First we introduce the operators which are needed
in the proving,
$$\eqalign{&O_k=\prod_i^{N_k}{(1+e(z_i(k),w_i(k)))\over 2} \, , \cr &
A_{k}^1 f(z_i(k))=\sum_P {(-1)^p \over N_k!} f(z_{Pi}(k)) \, , \cr  &
A_{k}^2 f(w_i(k))=\sum_P {(-1)^p \over N_k!} f(w_{Pi}(k)) \, , \cr}
\eqn\antizs$$
where $P$ are the permutations on $1,2,\ldots , N_k$.
Let us consider the simplest hierarchical wave function ($l=2$).
The wave function is now
$$\Psi=\int \prod dv_q
{F_1}^{p_1}G_1S_1 {(F_2)}^{-p_1\over 2p_1+1} G_2
{(\bar F_2)}^{{-p_1\over 2p_1+1}+p_2} {\bar G}_2  \, .
\eqn\aaz$$
$G_1$ satisfies FCC \aaq.
As an {\bf important } fact, we can prove that
$G_1 O_2(S_1G_2)$  also satisfies
FCC or the equation \aaq
$$E_{z_i}(G_1O_2(S_1G_2))=G_1O_2(S_1G_2) \, . \eqn\import$$
The formula \import is valid because
$$\eqalign{&\prod_{i<j}(z_i(1)-z_j(1)) (w_i(1)-w_j(1)) \times \cr &
O_2[\prod_{i,j}(z_i(1)-z_j(2)) (w_i(1)-w_j(2)) \times \cr &
\prod_{i<j}(z_i(2)-z_j(2)) (w_i(2)-w_j(2))] \cr} $$
satisfies FCC.
Now we shall show that
$$\int \prod dv_q {F_1}^{p_1}|(F_2)|^{-2p_1\over 2p_1+1} {(\bar F_2)}^{p_2}
G_1O_2(S_1G_2){\bar G}_2
\eqn\bba$$
is proportional to the wave function  $\Psi$ in \aaz. Because
$${F_1}^{p_1}|(F_2)|^{-2p_1\over 2p_1+1} {(\bar F_2)}^{p_2}$$
is the completely symmetric function of the coordinates
$z_i(2),w_i(2)$,  the operator $O_2$ acting on the function
$S_1G_2$ can be removed to act on the function ${\bar G}_2$
inside the integration.  So \bba is equal to
$$\int \prod dv_q
{F_1}^{p_1}|(F_2)|^{-2p_1\over 2p_1+1} {(\bar F_2)}^{p_2}
G_1S_1G_2   (O_2{\bar G}_2) \, .  \eqn\bbb$$
But $G_2$ is the anti-symmetric function
of the coordinates $z_i(2)$ and the anti-symmetric function
of the coordinates $w_i(2)$,  hence  \bbb will be equal to
$$\int \prod dv_q
{F_1}^{p_1}|(F_2)|^{-2p_1\over 2p_1+1} {(\bar F_2)}^{p_2}
G_1S_1G_2 (A_2^1A_2^2O_2){\bar G}_2 \, .\eqn\bbc$$
It can be shown that
$$(A_2^1A_2^2O_2){\bar G}_2$$
is proportional to ${\bar G}_2 $.  Hence
we can conclude that \bba is proportional to  $\Psi$ in \aaz.
But the formula \bba satisfies
FCC due to the identity \import ,
so  the wave  function \aaz also satisfies FCC.
For the  case of  the general hierarchical wave function,
we leave the proving to the next section.
We shall mention that, Moore and Read had discussed the
Halperin spin-singlet state from the point of view of
Conformal Field Theory [\GREAD].
They had also discussed ordinary spin polarized
hierarchical wave function by using Conformal Field Theory.
It will be interesting to see how
to obtain the above spin-singlet hierarchical wave function by
using the method developed in [\GREAD].

\chapter{OBTAINING THE HIERARCHICAL WAVE FUNCTION OF SFQHE FROM
MORE PHYSICAL POINT OF VIEW?}

The last section gives us an impression that we obtain
the hierarchical wave function \aar satisfying
FCC only by guess .
In fact, we did not get the
wave function \aar directly during this work.
In this section,  we shall  show our original reasoning
(based on the physical intuition) which
we used to obtain the hierarchical wave function.
The picture presented
in the following may be not right, but the final wave function obtained
in this picture is the same as that in the last section and we think that
it is worthwhile to include it here.
The picture presented in this section shall be called as pairing picture.
The pairing picture had firstly and extensively
been used by the authors in [\GREAD]
to construct spin-singlet state in FQHE, for example,
Halperin spin-singlet state, Haldane and Rezayi spin-singlet state.
Pfaffian state  at a filling factor
$1\over q$ with $q$ as even integer was also obtained
based on  pairing picture [\GREAD].
\par
If the  wave function $\Psi_{p_1+1,p_1+1,p_1}$
in the presence of quasi-particles is
$$\Psi^p_1=F^{p_1}_1(F_2)^{-p_1\over 2p_1+1}
G_1 O_2(S_1G_2) \, ,\eqn\bbd$$
then it will satisfy FCC \aaq  because
$G_1 O_2(S_1G_2)$   satisfies FCC.
{}From the wave function $\Psi^p_1$,  we shall suppose that
the Laughlin wave function of the quasi-particles  is
$$\Psi^p_2= ({\bar F}_2)^{{-p_1\over 2p_1+1}+p_2} O_2({\bar G}_2)
\, . \eqn\bbf$$
It is reasonable to assume that
the spin function of  the quasi-particles is
$$\prod {(\alpha_{z_i(2)}\beta_{w_i(2)}-\alpha_{w_i(2)}
\beta_{z_i(2)})\over{\sqrt 2}}  \, .\eqn\bbg$$
The spin function of  the quasi-particles given by \bbg
will insure that the excited state is the eigenstate of
$S^2$ with the eigenvalue being $0$.
The excitations of the Laughlin states look like {\it Skymion} excitations.
The Skymion excitation is specified by the coordinates
$z_i(2)$ and $w_i(2)$,  and it
is a bound state of the Laughlin quasi-particles at
$z_i(2)$ and $w_i(2)$.
This bound state looks like Cooper  pair in superconductivity.
We can demonstrate this point more clearly if we write the wave function
of the quasi-particles as
$$\Psi^{\prime}_2= |F|_2^{-p_1\over 2p_1+1} {\bar F}^{p_2}O_2({\bar G}_2)
\, , \eqn\bbh$$
which is related to the wave function in \bbf by a singular gauge
transformation. In the new wave function of the quasi-particles
(the wave function now is the spin function \bbg
multiplied by the  wave function \bbh),
when  $z_i(2)$ exchanges  with $w_i(2)$,
we shall get a minus sign and when exchange the  coordinates
$z_i(2)$ and $w_i(2)$ with the coordinates
$z_j(2)$ and $w_j(2)$, the sign of the wave function
remains unchanged. So in this gauge, the quasi-particle is
fermion and the bound state is  boson. Thus
it exactly looks likes the case of Cooper pairs in superconductivity.
\par
We can proceed to construct the next hierarchy in a similar way.
The quasi-particles in any hierarchy all
are bounded to pairs.
Then the general hierarchical wave function is
$$\Psi^p=\int \prod dv_q \Psi^p_1\Psi^p_2 \cdots \Psi^p_l \, , $$
with
$$\eqalign{&\Psi^p_1=F^{p_1}_1G_1(F_2)^{-p_1\over 2p_1+1}O_2(S_1G_2)
 \, ,\cr & \cdots   \,  ,\cr &
{\tilde \Psi^p_k}={(F_k)}^{s_k+p_k}
(F_{k+1})^{-(s_k+p_k)\over 2(s_k+p_k)+1} (O_kO_{k+1})(G_kS_kG_{k+1}) \, ,
\cr & \cdots   \,  , \cr &
{\tilde \Psi^p_l}={(F_l)}^{s_l+p_l}O_l(G_l) \, . \cr} \eqn\bbi$$
Because $\Psi^p_1$ satisfies FCC,
it is clear that $\Psi^p$ shall satisfy FCC.
\par
We surprisingly find
that $\Psi^p$ is proportional to $\Psi$ in \aar .  Take the simplest case,
$l=2$, the proving of the above statement is rather easy.
The wave function $\Psi^p$ is  then
$$\Psi^p =\int \prod dv_q (symmetric \, function \, of \, z_i(2),w_i(2))
O_2(S_1G_2) O_2(\bar G_2) \, , \eqn\bbj$$
where  $symmetric \, function \, of \, z_i(2),w_i(2)
={F_1}^{p_1}|(F_2)|^{-2p_1\over 2p_1+1} {(\bar F_2)}^{p_2}$.
Inside the integration,
$O_2(S_1G_2) O_2(\bar G_2)$ can be changed to
$(S_1G_2) (O_2O_2)(\bar G_2)=(S_1G_2) O_2(\bar G_2)$
due to $O_kO_k=O_k$.
The remaining proving can be found in the last section.
For the general case, we take $l=3$ as an example.
The wave function is now
$$\eqalign{\Psi^p =&\int \prod dv_q
(symmetric \, function \, of \, z_i(2),w_i(2))\cr &
(symmetric \, function \, of \, z_i(3),w_i(3)) \cr &
O_2(S_1G_2) (O_2O_3)(\bar G_2 \bar S_2 \bar G_3) O_3(G_3)\, . \cr}\eqn\bbk$$
Inside the integration, we can change
$$O_2(S_1G_2) (O_2O_3)(\bar G_2 S_2 G_3) O_3(G_3)$$
to $$(S_1G_2) (O_2(O_2O_3))(\bar G_2 \bar S_2 \bar G_3) O_3(G_3)$$
which is equal to
$$ S_1G_2 (O_2O_3)(\bar G_2 \bar S_2 \bar G_3) O_3(G_3) \, . \eqn\bbl$$
However because $S_1G_2$ is the
anti-symmetric function of the coordinates $z_i(1)$
and the anti-symmetric function of the coordinates $w_i(1)$,
so  \bbl  is equal to
$$S_1G_2 (A^1_2A^2_2O_2O_3)(\bar G_2 \bar S_2 \bar G_3)
O_3(G_3) \, . \eqn\bbm$$
It can be shown that
$A^1_2A^2_2O_2O_3(\bar G_2 \bar S_2 \bar G_3)$
is proportional  to $\bar G_2 O_3(\bar S_2 \bar G_3)$.
So \bbm is proportional to
$$ S_1G_2 \bar G_2 O_3(\bar S_2 \bar G_3) O_3(G_3) \, . \eqn\bbn$$
Using the same reasoning as the one
between the formula \bbk and
the formula \bbl ,  one can show that
the formula \bbn inside the integration
can be replaced by
$$ S_1G_2 \bar G_2 \bar S_2 \bar G_3 ((O_3O_3)(G_3))=
 S_1G_2 \bar G_2 \bar S_2 \bar G_3 O_3(G_3)  \, . \eqn\bbo$$
Because
$\bar S_2 \bar G_3$ is the anti-symmetric function of the coordinates
$z_i(3)$ and the anti-symmetric function of
the coordinates $w_i(3)$,  \bbo
turns out to be equal to
$$ S_1G_2 \bar G_2 \bar S_2 \bar G_3 O_3O_3(G_3)=
 S_1G_2 \bar G_2 \bar S_2 \bar G_3 A^1_3A^2_3O_3(G_3)  \, . \eqn\bbp$$
Due to $A^1_3A^2_3O_3(G_3)$ being
proportional to $G_3$, so \bbp is
proportional to
$$  S_1G_2 \bar G_2 \bar S_2 \bar G_3 G_3  \, . \eqn\bbq$$
Thus we finally conclude that $\Psi^p$ is
proportional to the corresponding $\Psi$ in the
last section when $l=3$. Actually they are
the same wave functions.  So $\Psi$ given by \aar also satisfies FCC.
If $l>3$,  the proving can follow the same way as that we did in the
case of $l=3$.
\par
Let us now give a summarization of the current section.
We have constructed the wave function
$\Psi^p$ which satisfies FCC.
We have also proven that the wave function $\Psi$
and  $\Psi^p$ are actually the same wave functions. Thus
automatically we show that the hierarchical wave function $\Psi$
in the last section satisfies FCC.

\chapter{PARTICLE-HOLE CONJUGATION}

Let us recall the  particle-hole conjugation in
the spin-polarized (un-layered) FQHE.
If there is a state of FQHE  with  the filling  $\nu ={1\over m}$,
then the filling of the
conjugated state is  $\nu^c =1-{1\over m}$.
The vacuum state of FQHE
is defined as there are no electrons in the lowest Landau level. So the
conjugate vacuum $\Omega$ is the state filled
with every orbital in the lowest
Landau level occupied [\FRACT],
$$\Omega (z_1,z_2, \cdots , z_{\phi +1})=
\prod_{i<j}^{\phi +1} d(z_i,z_j) \, ,  \eqn\bbr$$
where $\phi$ is the magnetic flux and ${\phi +1}$ is the number of the
orbital in the lowest Landau level.  If there is a state
$\Psi (z_1,z_2, \cdots , z_N)$, then the corresponding conjugate state is
$$\Psi_C=\int dv_1 \cdots dv_N \Omega (z_1,z_2, \cdots , z_{\phi +1})
\Psi^{\ast}(z_1,z_2, \cdots , z_N) \, .  \eqn\bbs$$
The Laughlin wave function of FQHE state
with  the filling $1\over m$ is
$\Psi(z_1,z_2, \cdots , z_N)=\prod _{i<j}^N d^m(z_i, z_j)$,
where we have  the relation $m(N-1)=\phi$.
Thus the conjugate state is
$$\eqalign{\Psi_C&=\int dv_1 \cdots dv_N \prod_{i<j}^{\phi +1} d(z_i,z_j)
\prod_{i<j}^N {\bar d}^m(z_i, z_j) \cr &
=\int dv_1 \cdots dv_N \prod_{N+1 \leq i<j \leq \phi +1} d(z_i,z_j) \cr &
\prod_{N+1 \leq i \leq \phi +1, 1 \leq i \leq N} d(z_i, z_j)  \phantom{=}
\prod_{1 \leq i \leq N} |d(z_i, z_j)|^2
{\bar d}^{m-1}(z_i, z_j) \, ,\cr} \eqn\bbu$$
The filling of this state is
$$\nu =1-{1\over m}={1\over \displaystyle 1+
{\strut 1\over \displaystyle m-1}}  \, ,\eqn\bbv$$
and the wave function \bbu actually belongs to the hierarchical wave function
constructed by Blok and Wen [\BW] (see also [\HALP] and [\LINN]).
We can show that, by using
the conjugation operation, the conjugate state of
the hierarchical wave function, of which  the filling is
$$\nu ={1\over \displaystyle p_1+
{\strut 1\over \displaystyle p_2+
{\strut 1\over \displaystyle \cdots +
{\strut 1\over \displaystyle p_l}}}}  \, ,\eqn\bbw$$
is given by another hierarchical state with the filling as
$$\nu_c=1-\nu={1\over \displaystyle 1+
{\strut 1\over \displaystyle p_1-1+
{\strut 1\over \displaystyle \cdots +
{\strut 1\over \displaystyle p_l}}}}  \, .\eqn\bbx$$
The conjugate vacuum state of the spin-singlet state
$\Omega_s$ now is
$$\Omega_s =\Phi_{1,1,0}(z_1,z_2,\cdots , z_{\phi+1};
w_1,w_2,\cdots , w_{\phi+1}) \, .\eqn\singlete$$
Now we consider the conjugate state of
$\Phi_{n+1,n+1,n}(z_1, z_2, \cdots , z_N; w_1, w_2, \cdots ,
w_N)$,
$$\Phi_c=\int \prod_{1\leq i \leq N}
dv_{z_i}dv_{w_i} \Phi^{\dagger}_{n+1,n+1,n}  \Omega_s \eqn\wangf$$
The filling of the state $\Phi_c$
is $2-{2\over 2n+1}={2\over {1+{1\over 2n}}}$,
and by explicit calculation,  $\Phi_c$
turns out to be the hierarchical state constructed in section $3$ or $4$
with $l=2, p_1=0, p_2=n$.
Generally, the conjugate wave function of
the hierarchical wave function
specified by  the parameters
$(p_1, p_2, \cdots , p_l)$ is the hierarchical wave function
specified by the parameters
$(p_1^{\prime}, p_2^{\prime},
\cdots , p_{l+1}^{\prime})=(0, p_1, p_2, \cdots , p_l)$.
The summation of the fillings of two states, which are
conjugate with each other,   is  always equal to $2$.
The above discussion offers
some kinds of checking to the hierarchical wave function constructed
in section $3$ or $4$. We finally remark that when $p_1=0,p_2=2, l=2$,
the filling is $8\over 5$, and this state is  conjugate to the
 state specified by the parameters $p_1=2, l=1$
with the filling as $2\over 5$.

\chapter{CONCLUSION}

We have constructed
the hierarchical wave function of SFQHE which satisfies FCC.
The particle-hole conjugation has been used to check the wave function.
We have also  discussed the spin of the quasi-particles
and the spin-statistics relation in some cases.
The hierarchical state of
SFQHE has also been discussed in [\REZY]. The relation between [\REZY] and
this work is not clear. There are some
other approaches to non-polarized FQHE,
for example, the effective Ginzburg-Landau theory approach
[\KANE,\STU]. Although the  Ginzburg-Landau theory of the
spin-polarized FQHE has  been well developed
([\BW,\CHERN,\ZHK,\RE,\WQ,etc.]
and also the review articles [\ROB,\ZEE]),
the Ginzburg-Landau theory of SFQHE has not been
much studied in the literature.
While we know how to implement the rotational invariance
condition on the wave function of SFQHE (on the sphere),
we do not know how to fully implement the rotational invariance
condition in the effective Ginzburg-Landau theory.
In the microscopic approach to SFQHE, the
rotational invariance condition on
the spin sector of the wave function
turns out to be Fock cyclic condition on the wave function.
The rotational invariance condition on the space sector will
give us a set of relations from which
we can derive the filling of the state. But how can we apply
the rotational invariance correctly and sufficiently
on the corresponding Ginzburg-Landau theory?
\par
We can also obtain the wave function of SFQHE
on the torus (the wave function of the spin-polarized
FQHE on the torus,  which has the  filling as $1\over m$
with $m$ being an odd integer,
has been constructed  in [\HR] and
the hierarchical wave function on the torus
has been constructed in [\LID]).
Although the rotational invariance is broken for the
space part of the wave function
(we have another important
invariance on the torus, that is
translational invariance), we can still
require that the wave function is the eigenstate of $S^2$
with the eigenvalue as $0$.
$\theta_a(\sum_i z_i)\theta_a(\sum_i w_i)
\prod_{i<j}\theta_3(z_i-z_j)\theta_3(w_i-w_j)$
is a solution to FCC and so it can be used to construct
the wave function on the torus. we need to use
Fay' trisecant identity [\MUN] to prove this Fock cyclic identity
on the torus. $\theta_a$ with $a=1,2,3,4$ are
the $\theta$ functions on the torus,
in which $\theta_3$ is the odd function [\MUN].

\chapter{ACKNOWLEDGEMENTS}
The author would like to thank Professors R. Iengo and Z.B. Su
for many useful discussions.

\refout
\end
\bye